\documentclass[a4paper,preprintnumbers,floatfix,superscriptaddress,pra,twocolumn,showpacs,notitlepage]{revtex4-1}

\usepackage[utf8]{inputenc}
\usepackage[T1]{fontenc}
\usepackage[sc,osf]{mathpazo}
\usepackage{amsmath, amsthm, amssymb,amsfonts}
\usepackage{graphicx}
\usepackage{dcolumn}
\usepackage{bm}
\usepackage{bbm}
\usepackage{hyperref}
\usepackage{mathtools}
\usepackage{comment}
\usepackage{color}

\def\1{\mathbf{1}}
\def\0{\mathbf{0}}

\DeclareMathOperator{\tr}{tr}
\DeclareMathOperator{\Tr}{Tr}

\def\p{\mathbf{p}}
\def\q{\mathbf{q}}

\newcommand{\beq}{\begin{equation}}
\newcommand{\eeq}{\end{equation}}
\newcommand{\bea}[1]{\begin{equation}\begin{array}{#1}}
\newcommand{\eea}{\end{array}\end{equation}}
\newcommand{\beqn}{\begin{eqnarray}}
\newcommand{\eeqn}{\end{eqnarray}}

\renewcommand{\rho}{\varrho}

\newcommand{\processnext}[1]{%
  \ifx\listfinish#1\empty\else\listact{#1}\expandafter\processnext\fi}

\newcommand{\figref}[1]{Fig.~\ref{#1}}

\newcommand{\appref}[1]{Appendix~\ref{#1}}

\newcommand{\ea}{\end{eqnarray}}
\newcommand{\ban}{\begin{eqnarray*}}
\newcommand{\ean}{\end{eqnarray*}}

{\begin{framed}\begin{small}}
{\end{small}\end{framed}}

\DeclareGraphicsExtensions{.pdf,.png,.jpg}

\makeindex

\begin{document}
\title{Device-Independent Tests of Entropy}
\date{\today}

\author{Rafael Chaves}
\affiliation{Institute for Physics \& FDM, University of Freiburg, 79104 Freiburg, Germany}
\affiliation{Institute for Theoretical Physics, University of Cologne, 50937 Cologne, Germany}
\author{Jonatan Bohr Brask}
\affiliation{D\'epartement de Physique Th\'eorique, Universit\'e de Gen\`eve, 1211 Gen\`eve, Switzerland}
\author{Nicolas Brunner}
\affiliation{D\'epartement de Physique Th\'eorique, Universit\'e de Gen\`eve, 1211 Gen\`eve, Switzerland}

\begin{abstract}
We show that the entropy of a message can be tested in a device-independent way. Specifically, we consider a prepare-and-measure scenario with classical or quantum communication, and develop two different methods for placing lower bounds on the communication entropy, given observable data. The first method is based on the framework of causal inference networks. The second technique, based on convex optimization, shows that quantum communication provides an advantage over classical, in the sense of requiring a lower entropy to reproduce given data. These ideas may serve as a basis for novel applications in device-independent quantum information processing. \end{abstract}

\maketitle

The development of device-independent (DI) quantum information processing has attracted growing attention recently. The main idea behind this new paradigm is to achieve quantum information tasks, and guarantee their secure implementation, based on observed data alone. Thus no assumption about the internal working of the devices used in the protocol is in principle required. Notably, realistic protocols for DI quantum cryptography \cite{acin07} and randomness generation \cite{colbeck,pironio} were presented, with proof-of-concept experiments for the second \cite{pironio,christensen}.

The strong security of DI protocols finds its origin in a more fundamental aspect of physics, namely the fact that certain physical quantities admit a model-independent description and can thus be certified in a DI way. The most striking example is Bell nonlocality \cite{bell,review}, which can be certified (via Bell inequality violation) by observing strong correlations between the results of distant measurements. Notably, this is possible in quantum theory, by performing well-chosen local measurements on distant entangled particles. More recently, it was shown that the dimension of an uncharacterized physical system (loosely speaking, the number of relevant degrees of freedom) can also be tested in a DI way \cite{brunner08,vertesi,wehner,gallego}. Conceptually, this allows us to study quantum theory inside a larger framework of physical theories , which already brought insight to quantum foundations \cite{barrett,vanDam,IC,brunner14}. From a more applied point of view, this allows for DI protocols and for black-box characterization of quantum systems \cite{mayers,vazirani,hendrych,*ahrens,rabello,moroder,yang}.

In this context, it is natural to ask whether there exist other physical quantities which admit a DI characterization. Here we show that this is the case by demonstrating that the entropy of a message can be tested in a DI way. Specifically, we present simple and efficient methods for placing lower bounds on the entropy of a classical (or quantum) communication based on observable data alone. We construct such ``entropy witnesses''  following two different approaches, first using the framework of causal inference networks \cite{pearl}, and second using convex optimization techniques. The first construction is very general, but usually gives suboptimal bounds. The second construction allows us to place tight bounds on the entropy of classical messages for given data. Moreover, it shows that quantum systems provide an advantage over classical ones, in the sense that they typically require lower entropy to reproduce a given set of data.

\begin{figure}[t]
\includegraphics[width=0.98\columnwidth]{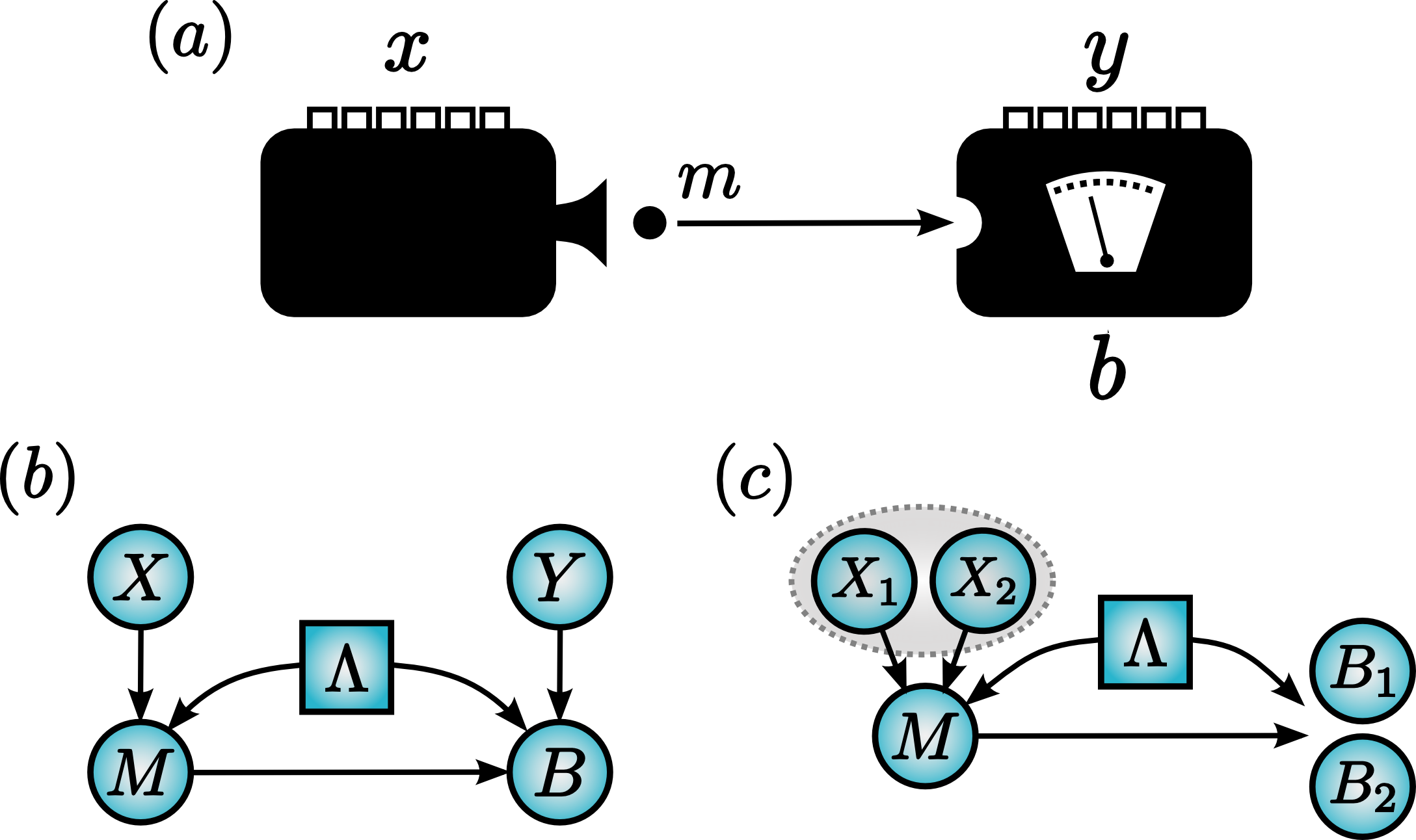}
\caption{Prepare-and-measure scenario. \textbf{(a)} Black-boxes representation. \textbf{(b)} Representation as a DAG.  \textbf{(c)} Finer description of the prepare-and-measure scenario where the number of measurements is explicitly taken in to account.}
\label{fig:PaM}
\end{figure}

\textit{Scenario.---}We consider the prepare-and-measure scenario depicted in \figref{fig:PaM}(a). It features two uncharacterized devices, hence represented by black-boxes: a preparation and a measurement device. Upon receiving input $x$ (chosen among $n$ possible settings), the preparation device sends a physical system to the measuring device. The state of the system may contain information about $x$. Upon receiving input $y$ (chosen among $l$ settings) and the physical system sent by the preparation device, the measuring device provides an outcome $b$ (with $k$ possible values). The experiment is thus fully characterized by the probability distribution $p(b\vert x,y)$. The inputs $x,y$ are chosen by the observer, from a distribution $p(x,y)$, which will be taken here to be uniform and independent, i.e. $p(x)=1/n$ and $p(y)=1/l$ (unless stated otherwise). A set of data $p(b\vert x,y)$ will also be represented using the vector notation ${\bf p}$; the $nlk$ components of ${\bf p}$ giving the probabilities $p(b\vert x,y)$.

Our main focus is the entropy of the mediating physical system, and our main goal will be to lower bound this entropy in a DI way, that is, based only on the observational data ${\bf p}$. We will consider both cases in which the mediating physical system is classical and quantum.

Let us first consider the quantum case. For each input $x$, the preparation device sends a quantum state $ \rho_x $ (in a Hilbert space of finite dimension $d$). We are interested in the von Neumann entropy of the average emitted state
\begin{equation}
S(\rho)= -tr(\rho \log \rho) \quad \text{where} \quad \rho=\sum_{x} p(x) \rho_x.
\end{equation}
Specifically we want to find the minimal $S(\rho)$ that is compatible with a given set of data, i.e. such that there exist states $ \rho_x$ and measurement operators $M_{b|y}$ (acting on $\mathbb{C}^d$) such that $p(b\vert x,y) = \tr(\rho_x M_{b|y})$. Note that in general we want to minimize $S(\rho)$ without any restriction on the dimension $d$.

In the case of classical systems, for each input $x$, a message $m \in \{ 0,...,d-1\}$ is sent with probability $p(m|x)$. The average message $M$ is given by the distribution $p(m)=\sum_{x}  p(m|x) p(x)$, with Shannon entropy
\begin{equation}
H(M) = - \sum_{m=0}^{d-1}  p(m) \log p(m).
\end{equation}
Again, for a given set of data, our goal is to find the minimal entropy compatible with the data, considering systems of arbitrary dimension $d$.

\textit{Entropy vs dimension.---} Since our goal is to derive DI bounds on the entropy without restricting the dimension our work is complementary to that of Gallego et al.~\cite{gallego}, where DI bounds on the dimension were derived. While the work of Ref.~\cite{gallego} derived DI lower bounds on worst case communication, our goal is to place DI lower bounds on the average communication.

More formally, Ref.~\cite{gallego} presented so-called (linear) dimension witnesses, of the form
\begin{equation}
V( {\bf p} ) =   {\bf v} \cdot {\bf p} =  \sum_{x,y,b} v_{xyb} p(b|x,y) \leq L_d,
\end{equation}
with (well-chosen) real coefficients $v_{xyb} $ and bound $L_d$. The inequality holds for any possible data generated with systems of dimension (at most) $d$. Hence if a given set of data ${\bf p}$ is found to violate a dimension witness, i.e. $V( {\bf p} )> L_d$, then this certifies the use of systems of dimension at least $d+1$.

In this work, we look for entropy witnesses, that is, functions $W$ which can be evaluated directly from the data ${\bf p}$ with the following properties. First, for any ${\bf p}$ requiring a limited entropy, say $H\leq H_0$, we have that
\begin{equation}
W( {\bf p} )  \leq L(H_0).
\end{equation}
Moreover, there should exist (at least) one set of data ${\bf p}_0$ such that $W( {\bf p}_0 )  > L(H_0)$, thus requiring entropy $H>H_0$. The problem is defined similarly for quantum systems, replacing the Shannon entropy with the von Neumann entropy.

Before discussing methods for constructing entropy witness, it is instructive to see that DI tests of entropy and dimension are in general completely different. Specifically, we show via a simple example, that certain sets of data may require the use of systems of arbitrarily large dimension $d$, but vanishing entropy.

Consider a prepare-and-measure scenario, and a strategy using classical systems of dimension $d+1$. We consider $n=d^2$ choices of preparations, and $l=n-1$ choices of measurements, each with binary outcome $b=\pm1$. Upon receiving input $x\leq d$, send message $m=x$; otherwise, send $m=0$. The entropy of the average message (with uniform choice of $x$) is found to be $H(M) = (2/d) \log(d) - (1- 1/d) \log(1- 1/d) $ which tends to zero when $n \rightarrow \infty$ (and hence $d\rightarrow \infty$). However, the corresponding set of data, ${\bf p}_0$, cannot be reproduced using classical systems of dimension $d$. This can be checked using a class of dimension witnesses \cite{gallego}:
\begin{equation}
\label{dimwit}
I_{n}( {\bf p} )=\sum_{y=1}^{n-1} E_{1y} +\sum_{x=2}^{n}\sum_{y=1}^{n+1-x} v_{xy} E_{xy} \leq L_d
\end{equation}
where $E_{xy} = \sum_{b= \pm 1} b \, p(b \vert x,y)$ and $v_{xy}=1 \text{ if } x+y \leq n \text{ and } -1 \text{ otherwise}$. For the above strategy, we obtain $I_n( {\bf p}_0 ) > L_d =n(n-3)/2 +2d -1$. Therefore, the data ${\bf p}_0$ requires dimension at least $d+1$ which diverges as $n \rightarrow \infty$, but has vanishingly small entropy in this limit.

\textit{Entropy Witnesses I.---}The above example shows that testing entropy or dimension are distinct problems. Thus new methods are required for constructing DI entropy witnesses. We first discuss a construction based on the entropic approach to causal inference \cite{pearl,quantumcausal,ChavesFritz2012,*FritzChaves2013,*Chaves2014,*Chaves2014b}. To the prepare-and-measure scenario of Fig. 1a, we associate a \emph{directed acyclic graph} (DAG) depicted in Fig. 1b. Each node of the graph represents a variable of the problem (inputs $X,Y$, output $B$, and message $M$), and the arrows indicate causal influence. Moreover, we allow the devices to act according to a common strategy, represented with an additional variable $\Lambda$ (taking values $\lambda$, with distribution $p(\lambda)$). We thus have that
\begin{equation}
\label{pbxy}
p(b\vert x,y)= \sum_{\lambda,m} p(b \vert y,m,\lambda)p(m\vert x,\lambda) p(\lambda).
\end{equation}

The key idea behind the entropic approach is the fact that the causal relationships of a given DAG are faithfully captured by linear equations in terms of entropies \cite{ChavesFritz2012,*FritzChaves2013,*Chaves2014,*Chaves2014b}. These relations, together with the so-called Shannon-type inequalities (valid for a collection of variables, regardless of any underlying causal structure), define a convex set (the entropic cone) which characterizes all the entropies compatible with a given causal structure. Note that for the quantum case, a similar analysis can be pursued, with the only notable difference that causal relations of the form \eqref{pbxy} must be replaced with data-processing inequalities; see \appref{app.entropicapproach} and Refs.~\cite{ChavesFritz2012,*FritzChaves2013,*Chaves2014,*Chaves2014b,quantumcausal} for more details.

Using the methods of \cite{ChavesFritz2012,*FritzChaves2013,*Chaves2014,*Chaves2014b,quantumcausal}, we characterized the facets of the entropic cone for the DAG of \figref{fig:PaM}(b). In the quantum case, the only non-trivial facet is given by
\begin{equation}
I(X:Y,B) \leq S(\rho) ,
\end{equation}
where $I(X:Y)=H(X)+H(Y)-H(X,Y)$ is the mutual information. Note that for the classical case, the Shannon entropy $H(M)$ replaces $S(\rho)$. The above inequality, which in fact follows directly from Holevo's bound \cite{holevo1973}, provides a simple and general bound for the entropy for given data, valid for an arbitrary number of preparations, measurements, and outcomes. However, this comes at the price of a very coarse-grained description of the data, and therefore will typically provide a poor lower bound on the entropy.

It is possible to obtain a finer description by accounting explicitly for the fact that the number of measurements $l$ is fixed. To do so, we replace the variables $Y,B$ with $l$ new variables $B_y$, and split the variable $X$ into $l$ separate variables $X=(X_1,\ldots,X_l)$; considering here $n=r^l$ for some integer $r$ \footnote{Note that the method also applies for arbitrary number of preparations $x$. Simply assign zero probability to all but $n$ of the possible inputs $(x_1,\ldots,x_l)$.}.

We first discuss the case of $l=2$ measurements. The corresponding DAG is illustrated in \figref{fig:PaM}(c).
Applying again the methods of Ref. \cite{ChavesFritz2012,*FritzChaves2013,*Chaves2014,*Chaves2014b}, we find a single non-trivial inequality (up to symmetries)
\begin{equation}\label{EW2}
\begin{split}
& I(X_1:B_1)+I(X_2:B_2)\\
&+I(X_1:X_2 \vert B_1)-I(X_1:X_2) \leq S(\rho).
\end{split}
\end{equation}
A general class of entropy witnesses can be obtained by extending the above inequality to the case of $l$ measurements (details in \appref{app.entropicwitness}):
\begin{equation}
\label{EWgen}
\begin{split}
& \sum_{i=1}^{l}I(X_i:B_i)+\sum_{i=2}^{l}I(X_1:X_i \vert B_i)  \\
&-\sum_{i=1}^{l}H(X_i)+H(X_1,\dots,X_l) \leq S(\rho) .
\end{split}
\end{equation}
These witnesses give relevant (although usually suboptimal) bounds on $S(\rho)$. For instance, we show in \appref{app.maxviolIn} that the maximal violation of the dimension witnesses $I_n(\p)$ (given in \eqref{dimwit}), which implies the use of systems of dimension $d=n$ \cite{gallego}, also implies maximal entropy, i.e. $ S(\rho) \geq \log n$.

We note that similar entropy witnesses can be derived for the case of classical communication. In fact, it suffices to replace $S(\rho)$ with $H(M)$ in \eqref{EW2} and \eqref{EWgen}. Note that \eqref{EWgen} is reminiscent of the principle of information causality \cite{IC}, but considering here a prepare-and-measure scenario \cite{brunner14,horodecki14}. That is, we consider classical correlations and quantum communication rather than quantum correlations and classical communication. Therefore, these witnesses cannot distinguish classical from quantum systems. More specifically, given a set of data, the classical and quantum bounds on the entropy will be the same, although this may not be the case in general, as we will see below.

To summarize, the entropic approach allows us to derive compact and versatile entropy witnesses, for scenarios involving an arbitrary number of preparations, measurements and outcomes. Moreover, the bounds obtained on the entropy are valid for systems of arbitrary dimension. Nevertheless, this approach has an important drawback, namely that the bounds we obtain will typically underestimate the minimum entropy actually required to produce a given set of data. The reason for this is that in general there exist many different sets of data giving rise to the same value of the witness \cite{entropicbell2}, e.g. the LHS of \eqref{EWgen}. The entropy bound will thus correspond to the lowest possible value $S(\rho)$ among these sets of data. This motivates us to investigate a different approach, which better exploits the structure of the data. We also note that for the witnesses above, we obtain the same entropy bound for classical and quantum systems. In the following, we will be able to distinguish them.

\textit{Entropy witnesses II.---} We will now discuss a method for placing bounds on the entropy using the entire set of data ${\bf p}$. This method can then be simplified to make use of only linear functions of the probabilities $p(b|x,y)$; in this case, we shall see that entropy witnesses can be directly constructed from dimension witnesses. This will allow us to show that, in the DI setting, quantum systems can outperform classical ones in terms of entropy.

Consider the case of classical communication. At first sight, one of the main difficulties is that we need to consider strategies involving messages of arbitrary dimension. However, notice that in the case of a finite number $n$ of preparations, we can focus on messages of dimension $d \leq n$ without loss of generality (see \appref{app.nsufficient}). It then follows that we have a finite number $D$ of deterministic strategies labeled by $\lambda$. For each strategy, the message $m$ is given by a deterministic function, $g_\lambda(x)$, and the output $b$ is given by a deterministic function $f_\lambda(y,m)$. Then, any set of data can be decomposed as convex combination over the deterministic strategies. More formally, we thus write $\p = {\bf A} \q$, where $\q$ is a $D$-dimensional vector with components $q_\lambda=p( \lambda)$ representing the probability to use strategy $\lambda$, and $\sum_{\lambda}q_\lambda =1$. The matrix ${\bf A}$, of size $nlk  \times D$, has elements $A_{(xyb),\lambda} = \delta_{b,f_\lambda(y,m)} \delta_{m,g_\lambda(x)}$.

The problem can thus be expressed as follows
\begin{equation}
\label{opt_prob}
\min H(M) \hspace{0.3cm} \text{s.t.} \hspace{0.3cm}  {\bf A} \q = \p, q_\lambda \geq 0 \text{  and  } \sum_{\lambda} q_\lambda=1 .
\end{equation}
where the minimization is taken over all possible convex combinations of deterministic strategies that reproduce $\p$. Notice that this set of possible convex decompositions of $\p$ forms a polytope $\mathbb{Q}$ (in the space of $\q$). Thus, although the objective function $H(M)$ is not linear in $\q$, this problem can be addressed by noting that $H(M)$ is concave in $\q$. It follows that the minimum of $H(M)$ will be obtained for one of the vertices of $\mathbb{Q}$.

The above procedure is analytical, and can therefore be applied for any given $\p$, in principle. However, it is computationally too demanding, even in the simplest cases, mainly due to the characterization of the polytope $\mathbb{Q}$. We thus further simplify the problem. First, we consider specific linear functions of the data $V(\p)$ (instead of the entire data $\p$). The first condition in \eqref{opt_prob} thus becomes $V({\bf A} \q) = V(\p)$. Moreover, we notice that this condition implies constraints on the distribution of the message $p(m)$, which can be characterized via a finite number of linear programs (see \appref{app.minHlinprog} for details).

\begin{figure}[t]
\includegraphics[width=0.95\columnwidth]{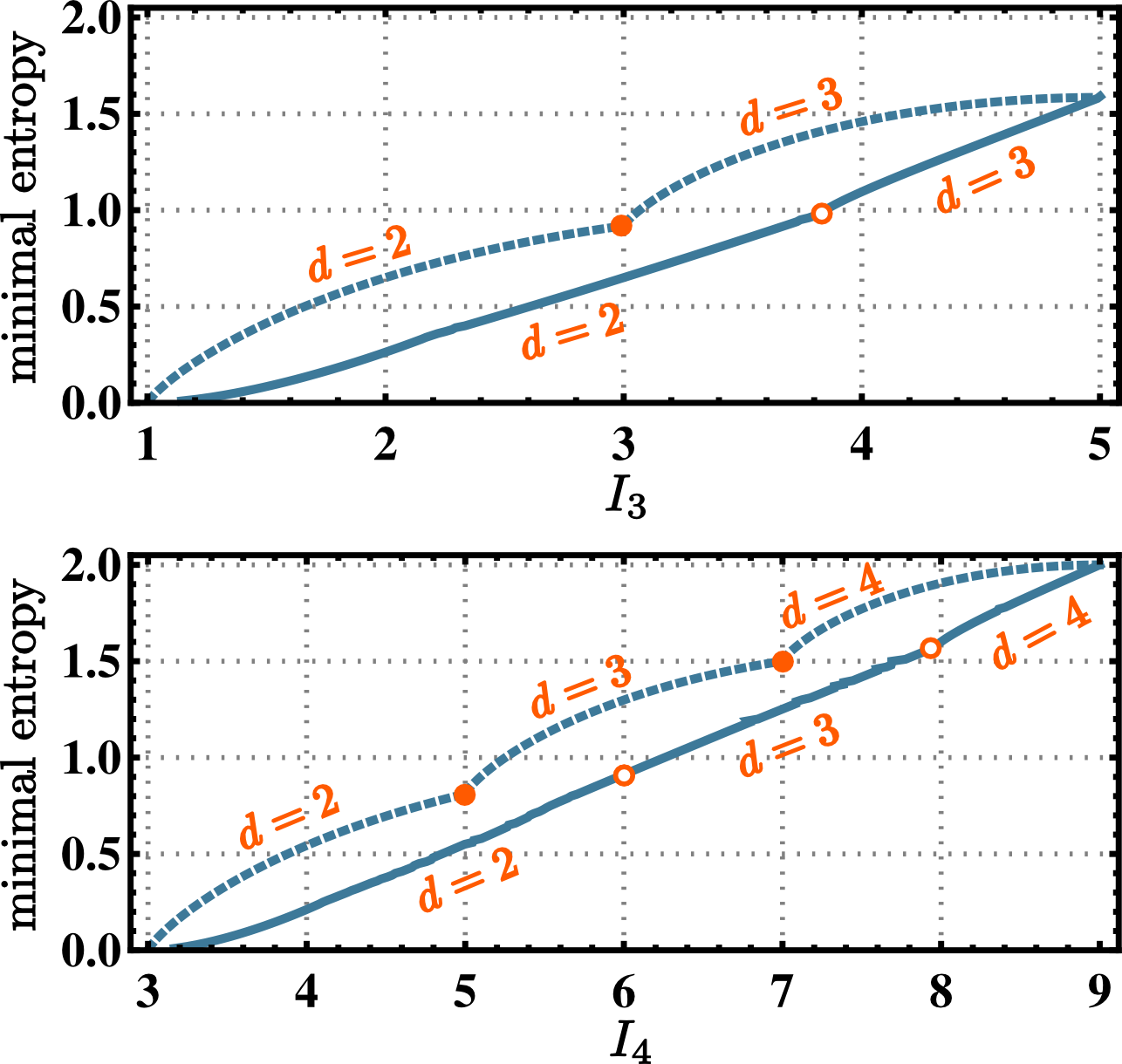}
\caption{Minimum values of $H(M)$ and $S(\rho)$ compatible with a given value of witnesses $I_3$ or $I_4$. Curves for classical (dotted) and quantum (solid) strategies are shown. The use of quantum strategies allow for a significant reduction in the communication entropy.}
\label{fig:I34}
\end{figure}

We apply this method to the linear dimension witnesses $I_n(\p)$ \eqref{dimwit} and illustrate it for $n=3,4$ (in \appref{app.minHlinprog} we also discuss the $2\rightarrow 1$ random access code). For each value of the witness, we obtain the minimum on the entropy $H(M)$ compatible with it. The result is shown in \figref{fig:I34}, and clearly shows that $\min H(M)$ is a non-trivial function of $I_n$. However, as we show next, $\min H(M)$ can be achieved with a very simple strategy. Consider that the value of $I_n$ lies in the range $ L_{d-1} \leq I_{n} \leq L_d$, that is, requires the use of $d$-dimensional states. Upon receiving input $x\leq d-1$, send message $m=x$; if $x = d$, send $m=d-1$ with probability $p=(L_d-I_n)/2$, and send $m=d$ with probability $(1-p)$; otherwise send $m=0$. The entropy of the average message is then
\begin{equation}
\label{minS}
 H(M) =  (d-2)\log{n}-\alpha\log{\alpha}-\beta\log{\beta},
\end{equation}
where $\alpha=(1-p)/n$ and $\beta=1-\alpha-(d-2)/n$ and which coincides (up to numerical precision) to the analytical bound for $\min H(M)$ for $I_3$, $I_4$, and $I_5$. Interestingly, this result shows that $\min H(M)$ requires only messages of minimal dimension; that is, for a given value of the witness $L_{d-1} < I_n(\p) \leq L_{d}$, systems of dimension $d$ are enough to achieve the lowest possible entropy. Another interesting feature is that no shared correlations between the preparation and measurement devices are needed. We also notice that, perhaps surprisingly, \eqref{minS} turns out to provide optimal entropy for all dimension witnesses that we have tested (see \appref{app.minHlinprog} further details). Whether this strategy is optimal for any dimension witness is an interesting open question. We highlight, nonetheless, that even if \eqref{minS} does not hold in general, it still provides a non-trivial upper bound on $\min H(M)$.

A relevant question is now to see if the use of quantum communication may help reducing the entropy. That is, for a given witness value, we ask what is the lowest possible entropy achievable using quantum systems. This is in general a difficult question, as we have no guarantee that using low-dimensional systems is optimal. Nevertheless, we can obtain upper bounds on $S(\rho)$ by considering low dimensional systems. We performed numerical optimization for quantum strategies involving systems up to dimension $d=4$ (see \appref{app.quantopt}). Results are presented in \figref{fig:I34}. Interestingly, the use of quantum systems allows for a clear reduction of the entropy (compared to classical messages) for basically any witness value. Whether the use of higher dimensional systems could help reduce $S(\rho)$ further is an interesting question.

\emph{Discussion.---}We have shown that the entropy of communication can be tested in a DI way, and presented two complementary methods tailored for this task. Our methods work for both classical and quantum communication, and the second method can be used to distinguish between classical and quantum systems for a given bound on the entropy.

Given the success of the DI approach for quantum information processing, it would be interesting to investigate potential applications based on the present work. While DI tests of dimension led to partially DI solutions for information tasks in the prepare-and-measure scenario \cite{PB11,lunghi15}, it would be relevant to explore the possibilities offered by DI entropy tests.

\emph{Acknowledgements.---}We thank Tam\'as V\'ertesi for useful discussions. We acknowledge financial support from the Excellence Initiative of the German Federal and State Governments (Grants ZUK 43 \& 81), the US Army Research Office under contracts W911NF-14-1-0098 and W911NF-14-1-0133 (Quantum Characterization, Verification, and Validation), the DFG (GRO 4334 \& SPP 1798), the Swiss National Science Foundation (grant PP00P2\_138917 and Starting Grant DIAQ), SEFRI (COST action MP1006), and the EU SIQS.

\bibliography{DI_entropy}

\appendix

\section{A brief review of the entropic approach to causal inference and its application in the prepare-and-measure scenario}
\label{app.entropicapproach}
The entropic approach for classical DAGs consists of three steps: (1) List all the Shannon type inequalities respected by a collection of $n$ variables, regardless of any underlying causal structure between them. (2) List the causal constraints that follow from a given causal structure. In terms of entropies these are linear constraints. (3) Marginalize the set of inequalities to the subspace of observable variables. Below we consider each of these steps and how they can be generalized to the quantum case, where some of the nodes in the DAG may represent quantum states. For more details see Refs.~\cite{ChavesFritz2012,*FritzChaves2013,*Chaves2014,*Chaves2014b,quantumcausal}.

\subsection{Step 1: Listing the Shannon type inequalities}
To understand these constraints, consider a collection of $n$ discrete random variables $X_1, \dots, X_n$ associated to some joint distribution $p(X_1, \dots,X_n)$. Let $X_T$ be the random vector $(X_i)_{i\in T}$ and denote by $H(T):=H(X_T)$ its Shannon entropy given by $H(X)=-\sum_{x}p(x)\log_2 p(x)$. Construct the associated entropy vector with $2^n$ real components, given by $h=(H(\emptyset),H(X_n),H(X_{n-1}),H(X_{n},X_{n-1}), \dots, H(X_1, \dots, X_n))$, to represent all the collections of entropies for $n$ variables. Not every vector in $\mathbb{R}^{2^n}$ will correspond to an entropy vector, as for example, entropies are positive quantities. The region of real vectors that correspond to entropies still lack an explicit description, however, an outer approximation to it is known, the so-called Shannon cone \cite{Yeung2008}.

The Shannon cone is characterized by two basic sets of linear constraints and a normalization constraint, the so-called Shannon-type inequalities. The first type are the monotonicity inequalities, for example, $H(X_1,X_2) \geq H(X_1)$, stating that the uncertainty about a set of variables should always be larger than or equal to the uncertainty about any subset of it. The second type of inequalities are given by the strong subadditivity condition which is equivalent to the positivity of the conditional mutual information. For example $I(X_1:X_2 \vert X_3)= H(X_1,X_3)+H(X_1,X_3)-H(X_1,X_2,X_3)-H(X_3) \geq 0$. Finally, the normalization constraint imposes $H(\emptyset)=0$.

\subsection{Step 2: Listing the causal constraints}
For an illustration, let us consider the DAG associated with the prepare-and-measure scenario, depicted in \figref{fig:PaM}(b).

Notice that in this causal structure we do not explicitly specify the numbers of different measurements or preparations. The causal constraints are encoded in the conditional independences implied by the causal structure. For instance, variables $X$ and $Y$ are not connected by arrows from one to the other nor is there a third variable connecting them. Thus, these variables should be statistically independent, which can be represented entropically via a linear relation $I(X:Y)=0$. In general, all the causal constraints following from a graph can be listed using the d-separation algorithm \cite{pearl}, but it is sufficient to use the so-called Markov decomposition. In the case of \figref{fig:PaM}(b) this states that
\begin{equation}
p(x,y,b)=\sum_{\lambda,m} p(b\vert y,m,\lambda)p(m\vert \lambda,x)p(x)p(y)p(\lambda).
\end{equation}
Using this decomposition, we can list all the relevant causal constraints for the DAG. These are
\begin{eqnarray}
& & H(X,Y,\Lambda)=H(X)+H(Y)+H(\Lambda), \\
& & H(M \vert X,\Lambda)=0  \\
& & H(B \vert Y,M,\Lambda)=0.
\end{eqnarray}
Notice that, without loss of generality, we imposed that $H(M \vert X,\Lambda)=0$ and $H(B \vert Y,M,\Lambda)=0$, basically saying that these variables are deterministic functions of their parents (any additional randomness can be absorbed in $\Lambda$).

\subsection{Step 3: Marginalization}

Given the description of the Shannon cone of $n$ variables plus the causal constraints, we are interested in its projection in the subspace containing only observable terms. This is achieved via a Fourier-Motzkin (FM) elimination \cite{Williams1986}. The final set of inequalities obtained via the FM elimination (and after eliminating over redundant inequalities) gives all the facets of the Shannon cone in the observable subspace. This set of inequalities consist of trivial and non-trivial ones. By non-trivial, we mean those inequalities that do not follow simply from the basic Shannon-type inequalities (monotonicity or strong subadditivity) but require the causal constraints to hold.

To illustrate, consider again the DAG of \figref{fig:PaM}(b). We marginalize over the variables that we do not have direct empirical access to, in this case $\Lambda$ and $M$. However, we still want to keep the term $H(M)$ as part of our description, because this is exactly the term we would like to bound from the observations of $X$, $Y$ and $B$. Proceeding with the marginalization step we find that the only non-trivial inequalities are
\begin{eqnarray}
\label{ineq_case1}
& & I(X:Y,B) \leq H(M), \\
& & I(X:Y)=0.
\end{eqnarray}

\subsection{Deriving entropic witnesses in the prepare-and-measure scenario}
\label{app.entropicwitness}
We now move on to the DAG in \figref{fig:PaM}(c). In this case, the causal constraints are given by
\begin{eqnarray}
& & H(X_1,X_2,\Lambda)=H(X_1,X_2)+H(\Lambda), \\
& & H(M \vert X_1,X_2,\Lambda)=0  \\
& & H(B_1,B_2 \vert M,\Lambda)=0.
\end{eqnarray}
Notice that we do not impose independence between the inputs, that is, $I(X_1:X_2) \neq 0$ in general. Performing FM elimination, we find that the only non-trivial inequalities are given by (up to permutations)
\begin{eqnarray}
\label{EDWX2trivial}
& & I(X_1,X2:B_1) \leq H(M), \\
\label{EDWX2}
& & I(X_1:B_1)+I(X_2:B_2)  \\ \nonumber
& & +I(X_1:X_2 \vert B_1) -I(X_1:X_2) \leq H(M) .
\end{eqnarray}
The first inequality is similar to what we have obtained above while the second inequality is the entropy witness described in the main text.

We notice that the same result holds true if we consider a modified DAG where the preparation and measurement devices are independent, i.e.~where the shared variable $\Lambda$ is split into independent variables $\Lambda_1$, $\Lambda_2$ connected to $M$ and to the $B$'s respectively with the new causal constraints
\begin{eqnarray}
& & H(X_1,X_2,\Lambda_1,\Lambda_2)=H(X_1,X_2)+H(\Lambda_1)+H(\Lambda_2), \\
& & H(M \vert X_1,X_2,\Lambda_1)=0  \\
& & H(B_1,B_2 \vert M,\Lambda_2)=0 .
\end{eqnarray}
This resembles the results discussed in the main text (c.f.~\eqref{minS}), where we have shown that the optimal strategy minimising the entropy for given values of $I_3$, $I_4$, and $I_5$ (and, we conjecture, $I_n$ in general) does not require shared correlations between the two devices.

Following the ideas in \cite{quantumcausal} we can prove that inequality \eqref{EDWX2} is also valid for a quantum message. The procedure is similar to the classical case though there are a few important differences. Because the message is quantum, we have to replace $H(M)$ by the von Neumann entropy $S(\rho)$. Another difference is that we cannot assign an entropy to $B$ and $\rho$ simultaneously. This is because for $B$ to assume a determined value we first need to a apply a completely positive, trace-preserving (CPTP) map on $\rho$ that in general will disturb $\rho$. Therefore, when constructing the set of inequalities and constraints we need to eliminate all those that contain $B$ and $\rho$ together, for example $S(\rho,B)$ and $S(\rho,X,B)$. A related problem is that one of the causal constraints valid in the classical case, $S(B\vert \rho,\Lambda)=0$, cannot be defined in the quantum case since it involves the term $S(\rho,\Lambda,B)$. The idea in \cite{quantumcausal} is to replace these causal constraints by corresponding data processing inequalities that are valid in quantum mechanics.

Following the approach in \cite{quantumcausal} we now prove that \eqref{EDWX2} and its generalization \eqref{EWgen} give valid bounds in the quantum case.
\begin{proof}
Rewrite the conditional mutual information appearing in \eqref{EWgen} as
\begin{equation}
I(X_1:X_i \vert B_i) = I(X_i:X_1, B_i)-I(X_i: B_i).
\end{equation}
Using this, the LHS of the inequality \eqref{EWgen} can be rewritten as
\begin{equation}
I(X_1:B_1)+\sum^{n}_{i=2}I(X_i:X_1,B_i)-\sum_{i=1}^{n}S(X_i)+S(X_1,\dots,X_n),
\end{equation}
This last expression can be upper bounded by
\begin{widetext}
\begin{eqnarray}
& \leq I(X_1:\Lambda,\rho) +\sum^{n}_{i=2}I(X_i:X_1,\Lambda,\rho)-\sum_{i=1}^{n}S(X_i)+S(X_1,\dots,X_n) \\
& =S(\Lambda,\rho)+(n-2)S(X_1,\Lambda,\rho) - \sum^{n}_{i=2} S(X_1,X_i,\Lambda,\rho)+S(X_1,\dots,X_n) \\
& \leq S(\Lambda,\rho) - S(X_1,\dots,X_{n},\Lambda,\rho)+S(X_1,\dots,X_n) \\
& \leq S(\Lambda,\rho) - S(X_1,\dots,X_{n},\Lambda)+S(X_1,\dots,X_n) \\
& = S(\Lambda,\rho) -S(\Lambda) \\
& \leq S(\rho)
\end{eqnarray}
\end{widetext}
which exactly gives \eqref{EWgen} as desired. In the above we have used: (i) data processing inequalities $I(X_i:B_i) \leq I(X_i:\Lambda,\rho) $ and $I(X_i:X_1,B_i) \leq I(X_i:X_1,\Lambda,\rho)$, (ii) the relation $-\sum^{n}_{i=2} S(X_1,X_i,\Lambda,\rho) \leq - S(X_1,\dots,X_{n},\Lambda,\rho)-(n-2)S(X_1,\Lambda,\rho) $, (iii) the monotonicity inequality $S(\rho\vert X_1,\dots,X_{n},\Lambda) \geq 0$, (iv) the independence relation $I(X_1,\dots, X_{n}:\Lambda)=0$, and (v) $S(\rho \vert \lambda) \leq S(\rho)$. Notice that we have used the von Neumann entropy $S$ for all terms. For the terms where all variables are purely classical (i.e.~that do not involve $\rho$), the von Neumann and Shannon entropies coincide, for example $S(X_i)=H(X_i)$.
\end{proof}

\section{Maximal violation of $I_n$ implies maximal entropy}
\label{app.maxviolIn}

As mentioned in the main text, inequality \eqref{EWgen} can be used to prove that a maximal violation of the dimension witness $I_n$, which implies message dimension $d=n$, also implies maximal entropy, i.e.~$S(\rho) \geq \log n$. We are interested in a scenario with $n$ preparations and $l=n-1$ measurements, with respective probabilities given by $p(x)=1/n$ and $p(y)=1/l$. Notice however, that in the construction of the inequality \eqref{EWgen} we have $l=n-1$ explicit variables $X_i$. To encode the probability $p(x)=1/n$ we consider each of the $X_i$ to be dichotomic variables and assign a joint probability distribution to them given by
\begin{equation}
\label{p1}
p\left(  x_1,\dots,x_{n-1} \right)  =\left\{
\begin{array}{ll}
\frac{1}{n} & , \,\, x_i=0 \,\, \forall i  \\
\frac{1}{n} & , \,\, x_i=1, x_{j \neq i}=0  \,\, \forall i \\
0 & \text{, otherwise}%
\end{array}
\right. .
\end{equation}
For example, in the case with $3$ preparations and $2$ measurements we have $p(0,0)=p(0,1)=p(1,0)=1/3$ and $p(1,1)=0$. Using the distribution $p(b\vert x,y)$ which achieves the maximum of $I_n$, and by direct calculation of the LHS of \eqref{EWgen}, we then find $ S(\rho) \geq \log n$. That is, to achieve the maximal violation of the dimension witness $I_n$, one needs maximal entropy, regardless of whether classical or quantum systems are used.

\section{Message dimension $n$ is sufficient}
\label{app.nsufficient}

In this section we prove that messages of dimension at most $n$ is required when minimising the entropy, where $n$ is the number of inputs for the preparation device. We will use the following terminology. A \textit{deterministic point} is an extremal point of the polytope in which the observed data $p(b|xy)$ lives. A \textit{deterministic strategy} is a recipe assigning deterministically a message to each given input for the preparation device and an output to each given message and input for the measurement device. Deterministic strategies are labelled by $\lambda$. The data can be decomposed as
\begin{align}
p(b|xy) & = \sum_{\lambda,m} p(b|my\lambda) p(m|x\lambda) p(\lambda) \\
& = \sum_\lambda \left(\sum_m \delta_{b,f_\lambda(m,y)} \delta_{m,g_\lambda(x)} \right) p(\lambda) \\
& = \sum_\lambda A_{bxy,\lambda} p(\lambda) \label{eq.pdetpointdecomp} .
\end{align}
Here $f_\lambda$, $g_\lambda$ are the deterministic functions specified by the strategy $\lambda$. The quantity $A_{bxy,\lambda}$ gives the deterministic point resulting from the strategy $\lambda$. In general there may be different deterministic strategies which result in the same deterministic point, i.e.~one can have $A_{bxy,\lambda}=A_{bxy,\lambda'}$ for different $\lambda$, $\lambda'$.

The probability for a certain message $m$ to occur is given by
\begin{align}
p(m) & = \sum_{\lambda,x} p(m|x\lambda) p(x) p(\lambda) \\
& = \sum_\lambda \left( \sum_x \delta_{m,g_\lambda(x)} p(x) \right) p(\lambda) \\
& = \sum_\lambda B_{m,\lambda} p(\lambda) , \label{eq.pmavg}
\end{align}
where $B_{m,\lambda}$ is the probability for $m$ given the strategy $\lambda$ averaged over the input distribution. Using this, the entropy of the message is
\begin{align}
H(M) & = -\sum_m p(m) \log(p(m)) \label{eq.entropy} \\
& = -\sum_m \left( \sum_\lambda B_{m,\lambda} p(\lambda) \right) \log\left(\sum_\lambda B_{m,\lambda} p(\lambda)\right) .
\end{align}
We are interested in what dimension is required for the message to achieve the minimum of this quantity, compatible with given observed data $p(b|xy)$.

We note that for fixed $\lambda$, the deterministic function $g_\lambda$ giving the message $m$ as a function of $x$ is fixed. Since there are $n$ inputs to the function, there can be at most $n$ different outputs. Therefore $m$ takes at most $n$ different values. Thus, for each deterministic strategy at most $n$ different values of $m$ occur. If all deterministic strategies make use of the same labels, then the total message dimension is at most $n$. However, it could in principle be that different strategies use different labels, such that the total message dimension is larger than $n$. This is not advantageous in terms of minimising the entropy though, as we now show.

For simplicity, consider just two deterministic strategies $\lambda_0$ and $\lambda_0'$. Denote the values of $m$ used in strategy $\lambda_0$ by $\mu_1,\ldots,\mu_n$ and let us assume that for strategy $\lambda_0'$ some of the labels are the same while some are different, e.g. $\mu_1,\dots,\mu_j,\mu_{j+1}',\ldots,\mu_n'$. Using \eqref{eq.pmavg} and \eqref{eq.entropy} we see that labels with $i\leq j$ will give rise to contributions to the entropy of the form
\begin{equation}
\left(B_{\mu_i,\lambda_0} + B_{\mu_i,\lambda_0'}\right) \log\left(B_{\mu_i,\lambda_0} + B_{\mu_i,\lambda_0'}\right) ,
\end{equation}
while the contributions from labels $\mu_i$, $\mu_i'$ with $i>j$ will be
\begin{equation}
B_{\mu_i,\lambda_0} \log\left(B_{\mu_i,\lambda_0}\right) + B_{\mu_i',\lambda_0'} \log\left(B_{\mu_i',\lambda_0'}\right).
\end{equation}
Now, for any two positive numbers $a$, $b$ with $a+b\leq 1$ one has that $(a+b)\log(a+b) \leq a\log(a)+b\log(b)$. It follows that, when minimizing the entropy, it is always advantageous to use the same set of labels in both strategies $\lambda_0$ and $\lambda_0'$. In general, one should use the same set of message labels in all the deterministic strategies needed to reproduce the data $p(b|xy)$, and hence a message of dimension at most $n$ is needed.

We note that, using concavity of the entropy, it is also possible to prove that there is no advantage in using several deterministic strategies for the same deterministic point. I.e.~it is optimal to take a single deterministic strategy for each deterministic point.

\section{Minimization of $H(m)$ as a linear program}
\label{app.minHlinprog}

As discussed in the main text, the minimum value of the entropy compatible with observed data $\p$ can be expressed as the following optimization problem (see also \cite{nonlocalcausal} for a statement of this problem in the context of Bell inequalities):
\begin{equation}
\label{opt_prob_app}
\min H(M) \hspace{0.3cm} \text{s.t.} \hspace{0.3cm}  {\bf A} \q = \p, \q_\lambda \geq 0 \text{  and  } \sum_{\lambda}\q_\lambda=1 .
\end{equation}
Given the concavity property of the entropy function, the minimum of $H(M)$ will be obtained at one of the vertices of the polytope defined by the linear constraints of the optimization problem \eqref{opt_prob_app}, which we denote $\mathbb{Q}$. However, the characterization of $\mathbb{Q}$ can be quite demanding computationally which leads us to introduce a simplified approach.

Notice that for evaluating $\min H(M)$ we only need to consider the probability distribution $p(m)=\sum_{\lambda,x} p(m\vert \lambda,x)p(\lambda)p(x)$ which, for a fixed value of $p(x)$, is therefore a linear function of the underlying hidden variable $\lambda$ (represented in \eqref{opt_prob_app} via the vector $\q$). The linear constraints in \eqref{opt_prob_app} will also imply linear constraints on $p(m)$. That is, the observable data defines a polytope $\mathbb{P}$ characterizing the probability $p(m)$ that is compatible with it. Therefore, to compute $H(M)$ we only need to consider the extremal points of $\mathbb{P}$. This significantly reduces the computational complexity of the problem and has allowed to us to consider the prepare-and-measure scenario with up to $n=5$ preparations and $l=4$ measurements.

To illustrate the general method for characterizing $\mathbb{P}$, in the following we will consider in details and without loss of generality the scenario $x\in\{0,1,2\}$, $m\in\{0,1,2\}$, $b\in\{0,1\}$ with all preparations equally likely, that is $p(x)=1/3$. Given the data $\p$ (or a linear function $V(\p)$ of it) the minimum and maximum values of $p(m)$ compatible with it can be found via \eqref{opt_prob_app}, where we simply replace the objective function $H(M)$ by $p(m)$.

\begin{figure}[t]
\includegraphics[width=0.99\columnwidth]{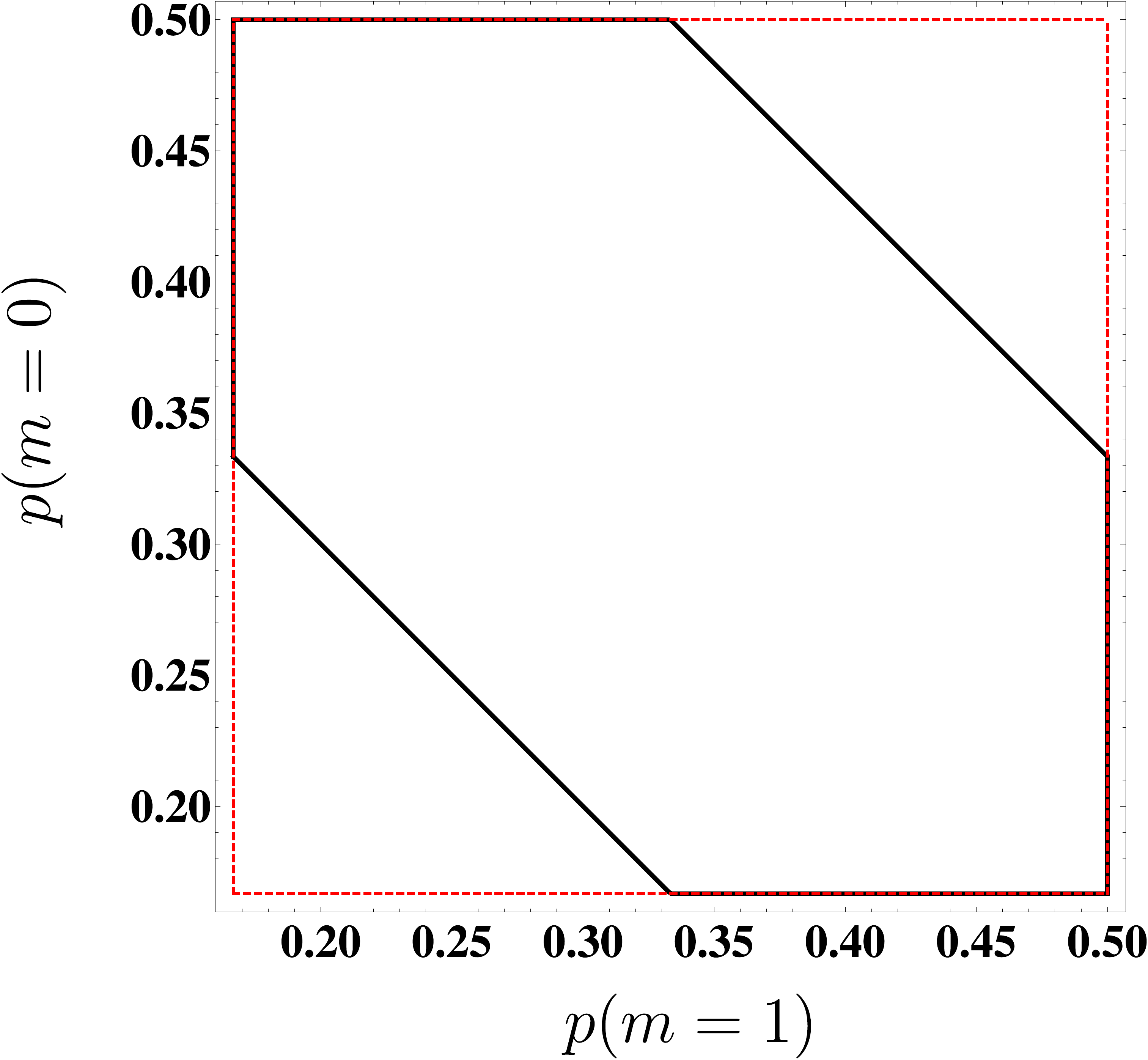}
\caption{In dashed red we see the polytopal region $1/6 \leq p(m) \leq 1/2$ and $\sum_{m} p(m)=1$. This is an outter approximation to the true polytopal region defined by the constraint $I_{3} =4$. The actual polytope can be found by solving a sequence of LPs and is shown in solid black.}
\label{fig:realpoly}
\end{figure}

For example, if we impose the constraint $I_{3}=4$ we find that $1/6 \leq p(m=0) \leq 1/2$. By symmetry the same holds true for $p(m=1)$ and $p(m=2)$ (since we are optimizing over all classical strategies, the labels we assign to $m$ are irrelevant). That is, under the constraint $I_{3}=4$ the minimum of $H(M)$ is restricted to be in the polytopal region defined by $1/6 \leq p(m) \leq 1/2$. Notice that by normalization we can write the entropy $H(m)$ as function of $p(m=0)$ and $p(m=1)$ alone, implying a $2$-dimensional polytopal region. The result is shown in \figref{fig:realpoly}. Also notice that the actual polytopal region implied by $I_{3}=4$ is smaller than (but contained) in $1/6 \leq p(m) \leq 1/2$. The reason is that further constraints, for example $ p(m=0)= 1/2$, will imply new constraints over $p(m=1)$, e.g.~$1/6 \leq p(m=1) \leq 1/3$. That is, this first polytope defines an outer approximation to the true polytope, and therefore provides only a lower bound (typically non-tight) on $H(m)$. The actual polytope $\mathbb{P}$ can be found by running a sequence of linear programs (LPs) as we explain next.

\begin{figure}[t]
\includegraphics[width=\columnwidth]{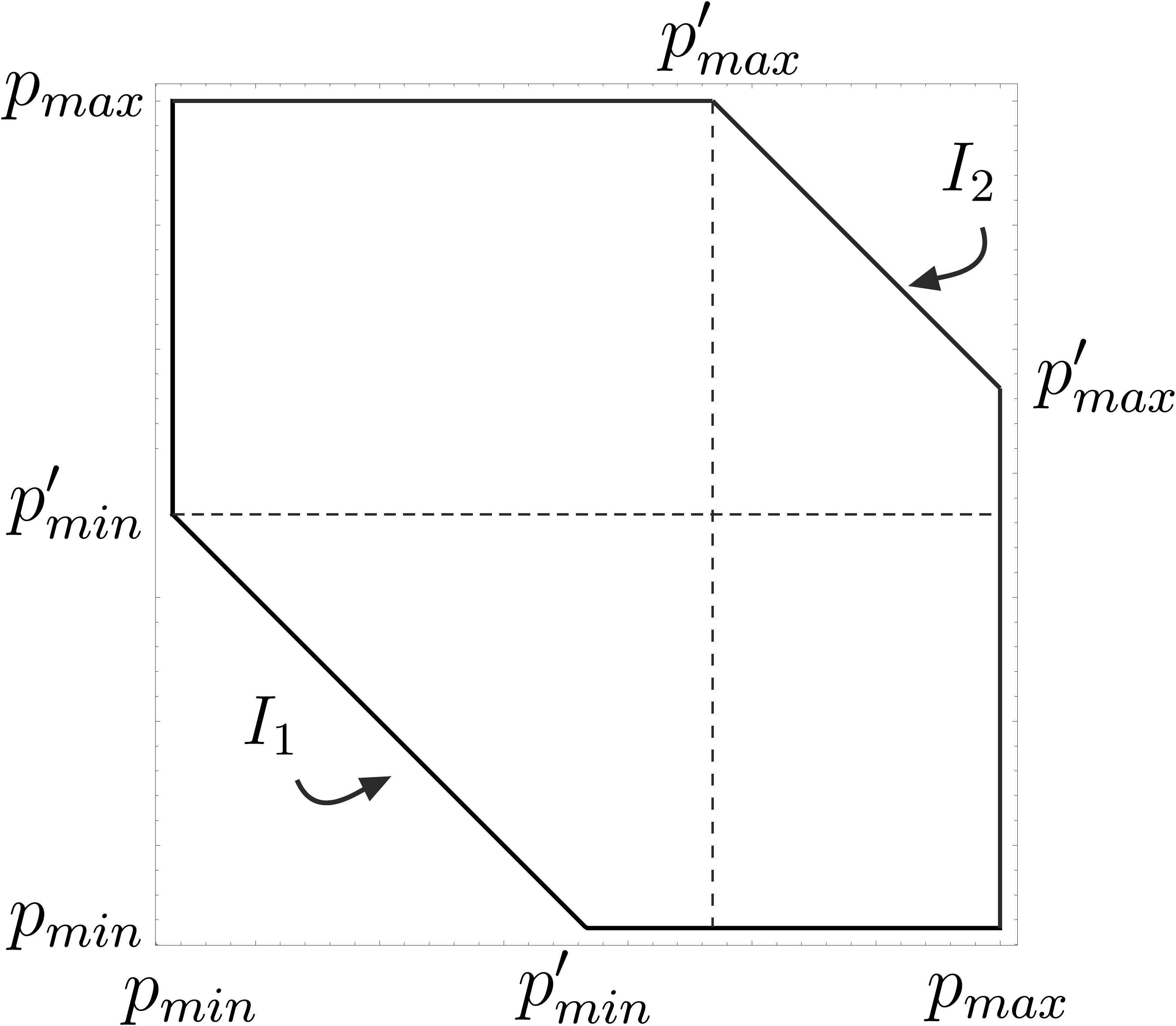}
\caption{Representation of the polytope $\mathbb{P}$ that can be found by solving a sequence of $4$ linear programs. The facets shown in figure correspond to $I_1 \rightarrow p(m=0)+p(m=1) \geq p_{min} +p^{\prime}_{min}$ and $I_2 \rightarrow p(m=0)+p(m=1) \leq p_{max} +p^{\prime}_{max}$. }
\label{fig:geo2}
\end{figure}

First one needs to run two LPs to find the bounds $p_{min} \leq p(m) \leq p_{max}$. Second, we need to find the maximum value $p^{\prime}_{max}$ of $p(m=1)$ under the constraint that $p(m=0)=p_{max}$ and the minimum value $p^{\prime}_{min}$ of $p(m=1)$ under the constraint $p(m=0)=p_{min}$. By symmetry, the value of $p^{\prime}_{max}$ and $p^{\prime}_{min}$ will be same if we reverse the roles of $p(m=0)$ and $p(m=1)$. From the fact that $p_{max}+p^{\prime}_{min}+p_{min} \leq 1$ and $p_{max} \geq p(m=2) = 1-p(m=0)-p(m=1)$ it follows that $p(m=0)+p(m=1) \geq p_{min} +p^{\prime}_{min}$. Similarly, from $p_{max}+p^{\prime}_{max}+p_{min}=1$ and $p(m=2) \geq p_{min} \rightarrow 1-p(m=0)-p(m=1) \geq p_{min}$ it follows that $p(m=0)+p(m=1) \leq p_{max} +p^{\prime}_{max}$. An illustration of this construction is shown in \figref{fig:geo2} and can be easily extended to higher dimensions.

In the main text we have used this procedure to compute $\min H(m)$ given values of the dimension witnesses $I_3$ and $I_4$, obtaining the general relation \eqref{minS} that we conjecture to be true for any $I_n$. Furthermore, the relation \eqref{minS} can be seen to hold for other classes of dimension witnesses. To illustrate this point, we consider the following inequality in the scenario with $n=4$ preparations and $l=2$ measurements \cite{PB11}
\begin{equation}
\label{DW42}
R_{4}=E_{11}+E_{12}+E_{21}-E_{22}-E_{31}+E_{32}-E_{41}-E_{42} \leq L^R_d .
\end{equation}
where, $L^R_d=2d$ for $d\geq 2$ and $L^R_d=0$ for $d=1$. The quantity $R_4$ quantifies the score in a $2\rightarrow 1$ random access code (RAC) game. In a RAC game one party (corresponding to our preparation device) receives a string of bits and then transmits a message to a second party (corresponding to our measurement device). Given the message and an index labelling one of the input bits, the second party must produce a binary outcome equal to that bit. $R_4$ corresponds to the case where the preparation device receives 2 bits and the measurement device receives 1 bit and produces a binary outcome.

\begin{figure}[t]
\includegraphics[width=0.99\columnwidth]{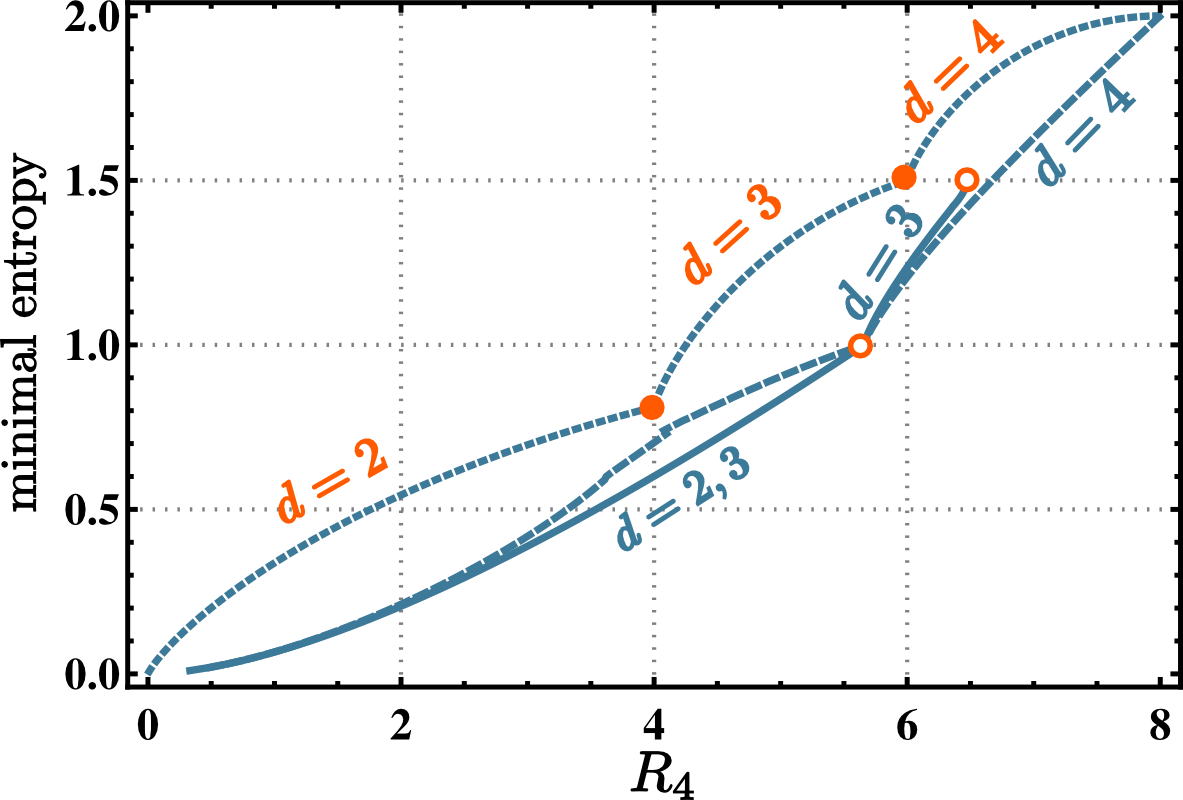}
\caption{Minimum values of the entropies $H(M)$, $S(\rho)$ for classical (dotted) and quantum (solid, dashed) strategies compatible with a given value of $R_4$. Solid and open circles indicate the points of maximal witness value achievable with the indicated dimension for classical and quantum strategies respectively. The quantum curves show the optimisation for qubits and qutrits (solid) and for real quqarts (dashed).}
\label{fig:R4}
\end{figure}

A crucial difference between $R_4$ and the class $I_n$ resides on the fact that for $I_n \leq n(n-3)/2+1$, $1$-dimensional messages (and therefore with zero entropy) are enough to reproduce the data. In contrast, for any $R_4 \neq 0$ we need at least $2$-dimensional systems. We have followed the same steps as for $I_n$ and obtained the classical curve in \figref{fig:R4}. This result is perfectly fitted by the same expression \eqref{minS} as for the $I_n$ class in the region $ 4 \leq L^R_d \leq 8$. However, it fails for $L^R_d \leq 4$. As discussed above this is exactly the region where $R_4$ and the $I_n$ class display a very different qualitative behaviour. Therefore such a difference for $d=2$ should come as no surprise. In the region $ L^R_d \leq 4$, the minimum entropy is described by $\min H(M)= H_\text{bin}((1/16)R_4)$, where $H_\text{bin}(x)=-x\log_2 x -(1-x)\log_2 (1-x)$ stands for the binary entropy.

As for $I_3$, $I_4$ we have also performed a numerical optimisation for quantum strategies, as explained below. The results are shown in \figref{fig:R4}. We see that, as for $I_3$, $I_4$, quantum strategies allow a significant reduction in entropy, i.e.~in average communication. The optimisation for quantum strategies was performed for qubits, qutrits, and for real ququarts. Interestingly, unlike for $I_3$, $I_4$, our results indicate that complex phases are necessary to reach the optimum for the RAC. The real ququart curve does not recover the results for qubits and qutrits in the region achievable with qubits (note though that the numerics are not completely stable in this region as may be seen from the plot). This suggests that real quqarts may also not be optimal above this region. At the same time, while in the qubit region we find no advantage for qutrits, we do find an advantage of ququarts over qutrits above the qubit region.

\section{Upper bounding the maximum dimension}
\label{app.boundim}

As discussed in the main text, for all the cases considered we observe that if given data can be reproduced with a classical message of dimension $d$, the minimum entropy $H(M)$ is also achieved with this dimension. In the following we give a geometric explanation for this effect. We consider without loss of generality the case where the data can be reproduced with $d=2$ and show that allowing for $d=3$ cannot lead to a smaller $H(M)$.

\begin{figure}[t]
\includegraphics[width=\columnwidth]{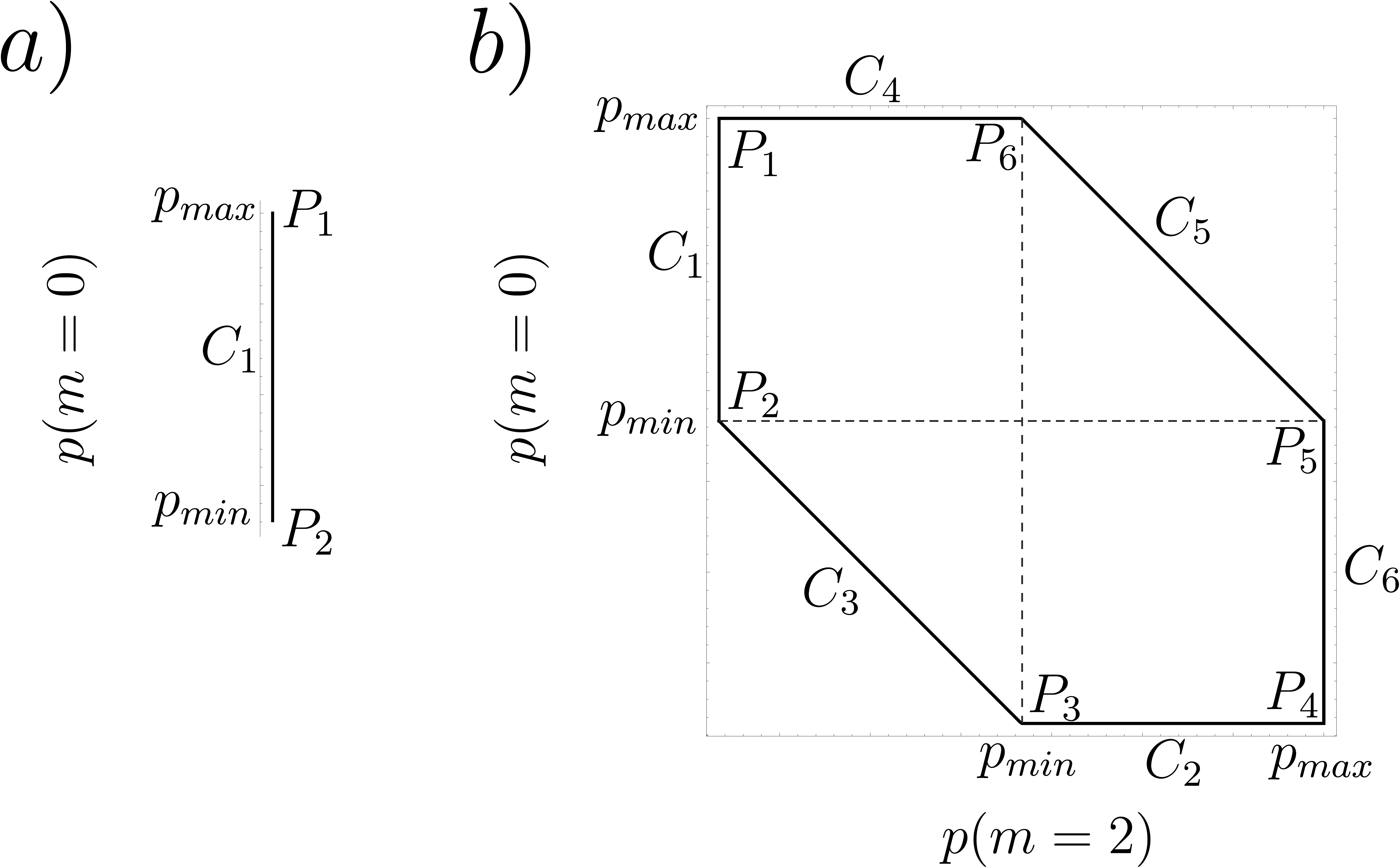}
\caption{Polytopes $\mathbb{P}_1$ and $\mathbb{P}_2$.}
\label{fig:geo1}
\end{figure}

In the $d=2$ case, because of the normalization constraint $p(m=0)+p(m=1)=1$ the polytope $\mathbb{P}_1$ can be represented as a $1$-dimensional object simply given by $p_{min} \leq p(m=0) \leq p_{max}$ (see \figref{fig:geo1}(a)). By concavity it follows that the minimum entropy over this set is $\min H(M)=\min \left[ H_{\text{bin}}(p_{min}),H_{\text{bin}}(p_{max}) \right]$. Consider now that we allow for $d=3$ leading to a $2$-dimensional polytope $\mathbb{P}_2$ that we parametrize as a function of $p(m=0)$ and $p(m=2)$. To show that this extra dimension cannot improve $H(M)$ it is sufficient to give an (in principle) outer approximation of $\mathbb{P}_2$ and show that $H(M)$ on all the extremal points of this set is larger than or equal to $\min H(M)$.

The polytope $\mathbb{P}_2$ is characterized by the following constraints (see \figref{fig:geo1}(b))
\begin{eqnarray}
& & C_1: 0 \leq p(m=2) , \\
& & C_2: 0 \leq p(m=0), \\
& & C_3:p(m=0) +p(m=2) \geq p_{min}, \\
& & C_4:p(m=0) \leq p_{max}, \\
& & C_5:p(m=0) + p(m=2) \leq p_{max}+p_{min}, \\
& & C_6: p(m=2) \leq p_{max}.
\end{eqnarray}
Constraints $C_1$, $C_2$, $C_5$ and $C_6$ trivially follow; using $p_{max}+p_{min} = 1$ we can easily prove $C_4$ and $C_5$ from $1-p(m=0)-p(m=2) \leq p_{max}$ and $p(m=0) + p(m=2) \leq 1 $, respectively. Therefore, polytope $\mathbb{P}_2$ is characterized by six extremal points, two of which are also extremal points of $\mathbb{P}_1$. Defining $H(\alpha,\beta)=-\alpha \log \alpha -\beta \log \beta -(1-\alpha -\beta) \log (1-\alpha -\beta)$, the entropy of the extremal points $P_1$ to $P_6$ are given, respectively, by $H_1=H(p_{max},0)$, $H_2=H(p_{min},0)$, $H_3=H(0,p_{min})$, $H_4=H(0,p_{max})$, $H_5=H(p_{max},p_{min})$ and $H_6=H(p_{min},p_{max})$. It follows that $H_1=H_3$, $H_2=H_4$ and $H_5=H_6$. Therefore, to prove that this extra dimension cannot improve $H(M)$ we only have to prove that $H_5 \geq H_1$ and $H_5 \geq H_2$. The inequality $H_5 \geq H_1$ is equivalent to
\begin{eqnarray}
\nonumber
& & -p_{min} \log p_{min} -(1-p_{min}-p_{max}) \log (1-p_{min}-p_{max})  \\
& &  \geq -(1-p_{max}) \log (1-p_{max})
\end{eqnarray}
that is trivially true since $p_{max}+p_{min} = 1$. Similarly one can prove that $H_5 \geq H_2$ which concludes the proof.

\section{Bounding the entropy for quantum strategies}
\label{app.quantopt}

In \figref{fig:I34} of the main text we show curves for the quantum entropy compatible with given values of the witnesses $I_3$, $I_4$. These curves are obtained by numerical optimisation and should be understood as upper bounds on the minimal quantum entropy.

Specifically, the curves are obtained by maximising the witness value while restricting the entropy $S(\rho)\leq s$ and then increasing $s$ from zero until the maximal witness value is reached. A priori one might expect that the optimisation should be performed over both preparations and measurements. However, it is only necessary to optimise over states because of the following observation. For a given choice of preparations $\rho_1,\ldots,\rho_n$ and measurements $M_1,\ldots,M_{n-1}$ the expected quantum value of the witness $I_n$ is given by (c.f.~\eqref{dimwit})
\begin{align}
I_n^q & = \sum_{y=1}^{n-1} \Tr [\rho_1 M_y] + \sum_{x=2}^n \sum_{y=1}^{n+1-x} \nu_{xy} \Tr[\rho_x M_y] \nonumber \\
& = \sum_{y=1}^{n-1} \Tr[(\rho_1 + \sum_{x=2}^{n+1-y}\nu_{xy}\rho_x) M_y] \nonumber \\
& = \sum_{y=1}^{n-1} \Tr[\rho_y' M_y] ,
\end{align}
with
\begin{equation}
\rho_y' = \rho_1 + \sum_{x=2}^{n+1-y}\nu_{xy}\rho_x .
\end{equation}
The observables $M_y$ are binary, so they are hermitian operators with eigenvalues $\pm 1$. Since the states $\rho_x$ are hermetian so are the sums of them $\rho_y'$. The maximal value of $I_n^q$ is then attained by choosing $M_y$ to be diagonal in the same basis as $\rho_y'$ with eigenvalues $\pm 1$ on the subspaces where $\rho_y'$ has positive and negative eigenvalues respectively. The maximum is thus equal to
\begin{equation}
\label{eq.Inqeigenvals}
I_n^q = \sum_{y=1}^{n-1} \sum_k |\lambda_{yk}| ,
\end{equation}
where $\lambda_{yk}$ are the eigenvalues of $\rho_y'$. To obtain the curves in \figref{fig:I34} we pick a dimension, e.g.~qubits, qutrits, or ququarts, we parametrise the states $\rho_x$, and we numerically maximise \eqref{eq.Inqeigenvals} subject to $S(\rho)\leq s$, where $\rho = \sum \rho_x / n$ is the average state assuming uniform inputs. The optimisation is implemented using NMaximize in \textit{Mathematica}.

For the witness $I_3$ we have performed the optimisation using fully parametrised qubits and qutrits, and using real ququarts (i.e.~paramtrisation without complex phases). We find that, for the range of values of $I_3$ which can be achieved by qubits, neither qutrits nor ququarts provide any advantage in terms of lowering the entropy (and in fact real qubits and real qutrits are sufficient).

For $I_4$ we have performed the optimisation for fully parametrised qubits, and for real qutrits and ququarts. Again we find that in the range achievable by qutrits, ququarts provide no advantage and in most of the range achievable by qubits, qutrits and ququarts provide no advantage. As before, real qubits perform the same as when phases are included, indicating that this may also be true for higher dimensions. We do, however, observe a small advantage of qutrits and ququarts in a narrow part of the qubit region, from $I_4\approx 5.52$ to $I_4=6$. The minimal entropy achieved by qubits in this region is a few percent larger than for qutrits and ququarts according to our numerical results. We believe this is due to suboptimal performance of the optimisation algorithm, although we have found no better point despite extensive testing.

\end{document}